# Direct observation of thermal hysteresis in the molecular dynamics of barocaloric neopentyl glycol

*Frederic Rendell-Bhatti[1*], Markus Appel[2], Connor S. Inglis[1], Melony Dilshad[3], Neha Mehta[4], Jonathan Radcliffe[4], Xavier Moya[3], Donald A. MacLaren[1] and David Boldrin[1]*

[1]SUPA, School of Physics and Astronomy, University of Glasgow, Glasgow G12 8QQ, United Kingdom

[2]Institut Laue Langevin, 71 Avenue des Martyr, 38000 Grenoble, France

[3]Department of Materials Science & Metallurgy, University of Cambridge, Cambridge, CB3 0FS, United Kingdom

[4]School of Chemical Engineering, University of Birmingham, Birmingham B15 2TT, United Kingdom

*Corresponding author email: fred.rendell@glasgow.ac.uk*

## Abstract

Barocalorics (BCs) are emerging as promising alternatives to vapour-phase refrigerants, which are problematic as they exacerbate climate change when they inevitably leak into the atmosphere. However, the commercialisation of BC refrigerants is significantly hindered by hysteresis in the solid-solid phase transition that would be exploited in a refrigeration cycle. Here, we provide new insight into the hysteresis that is a critical step towards the rational design of viable BCs. By studying the benchmark BC plastic crystal, neopentyl glycol (NPG), we observe directly the liberation of the hydroxyl rotational modes that unlock the hydrogen bond network, distinguishing for the first time the molecular reorientation and hydroxymethyl rotational modes. We showcase the use high-resolution inelastic fixed-window scans in combination with quasielastic neutron scattering (QENS) measurements to build a comprehensive microscopic understanding of the NPG phase transition, directly tracking the molecular dynamics of the phase transition. Hysteresis previously observed in calorimetric studies of NPG is now observed directly as hysteresis in molecular rotational modes, and hence in the formation and disruption of hydrogen bonding. Furthermore, by tracking the thermal activation of three main reorientation modes, we suggest that their fractional excitations may resolve an outstanding discrepancy between measured and calculated entropy change. These results allow for direct study of the molecular dynamics that govern the thermal hysteresis of small molecule energy materials. They will be broadly applicable, as many promising BC material families possess first-order transitions involving molecular reorientations.

# Introduction

Heat pumps, powered by renewable electricity, present a low carbon alternative for both heating [1], and cooling applications [2]. However, conventional heating, ventilation, and air conditioning (HVAC) systems employ gaseous refrigerants that leak into the atmosphere and have appreciable global warming potential (GWP)[3]. This problem will become a pressing concern as annual heat pump installations are projected to increase eleven-fold by 2028 in the UK alone[4]. As such, we require new refrigerant materials that are efficient, low-cost, safe and environmentally sustainable[5]. Solid-state barocaloric (BC) materials exhibit significant latent heat associated with solid-solid (S-S) phase transitions that can be driven by pressure, and therefore have the potential to revolutionise new and existing HVAC technologies through their use as working bodies[6–10]. Being solids, these materials have negligible GWP and have the potential for even higher efficiency than vapour refrigerants[11]. A promising class of BC materials includes the so-called molecular, or plastic, crystals (PCs), particularly neopentyl glycol (NPG), which exhibits a 'colossal' BC effect[12,13,17]. However, NPG is not itself viable for use in commercial HVAC systems. It exhibits substantial hysteresis when thermally cycled, necessitating high pressures to achieve the large reversible entropy changes needed for an efficient refrigeration cycle. Recent efforts have been made to both understand[14–17] and influence[18,19] the BC effect in NPG, which has become a benchmark for PC BCs. However, there is still significant opportunity for further development of PCs, which will rely on a better understanding of the microscopic mechanisms behind hysteresis in the S-S phase transition.

NPG, $(CH_3)_2C(CH_2OH)_2$, is a roughly spherical molecule consisting of two methyl (-$CH_3$) and two hydroxymethyl (-$CH_2OH$) groups that can all rotate about bonds to a central carbon atom. Below its melting temperature, $T_m$ = 400 K, NPG molecules adopt two distinct crystal structures. On heating at ambient pressures, NPG undergoes a S-S phase transition at $T_0$ = 314 K, where hydrogen bonds between molecules are broken and the structure changes from a monoclinic ordered crystal (OC) phase to a face-centered-cubic (fcc) phase (see Supporting Information Fig. S1). The latter is known as a disordered plastic crystal (PC) because molecular rotations are active and only the molecular centre of mass has a well-defined location[20,21]. The S-S, order-disorder phase transition is accompanied by large entropy ($\Delta S$ ~ 390 J K$^{-1}$ kg$^{-1}$) and volume ($\Delta V$ = 5 %) changes [13], and under adiabatic conditions the latter allows material temperature changes to be driven by pressure, which is the BC effect. The phase transition is consistent with both the activation of rotational dynamical modes and a disruption of the hydrogen bond network that locks molecular orientations in the OC phase. The behaviour of these rotational modes, and their thermal activation, offers direct insight into the mechanisms underpinning BC performance. Our ambition is to use this insight to tailor the microscopic mechanisms of the phase transition and thereby reduce thermal hysteresis during refrigeration cycles.

# Experimental Section

### QENS data acquisition

Powder samples of NPG (99% purity) were purchased from Sigma-Aldrich and used as received. Neutron scattering experiments were performed on the IN5 and IN16B spectrometers at the Institut Laue–Langevin (ILL), France. IN5 is a disk chopper time-of-flight spectrometer and IN16B is a backscattering spectrometer. On IN5 a wavelength of 6 Å was used, yielding a FWHM energy resolution of 60 µeV, dynamic range of ± 1.3 meV and $Q$-range of 0.10 to 1.89 Å$^{-1}$ in this experiment. IN16B was used in standard configuration with strained Si111 Doppler monochromator and analysers yielding a FWHM energy resolution of 0.75 µeV, dynamic range of ± 0.028 meV and $Q$-range of 0.19 to 1.89 Å$^{-1}$. QENS measurements were obtained using scan times of 2 hrs on IN16B and 30 mins on IN5. All QENS measurements were obtained on heating. FWS measurements were performed during a temperature ramp of approximately 0.5 K min$^{-1}$ with alternating acquisitions of elastic (30 s) and inelastic intensity at 3 µeV energy transfer (90 s). Approximately 0.5g of sample was loaded into an aluminium can with annular geometry for measurements on both IN16B and IN5. Empty can and vanadium standard measurements were acquired using the same sample geometry and used to correct the NPG data. Datasets from each spectrometer were treated independently from each other, using separate standard measurements for correction. We found that combining instrument QENS datasets (IN16B and IN5) gave the same results as treating the datasets separately. This is because the linewidths of Modes 2 and 3 were within the instrument resolution of IN5 (making them undetectable) and the linewidth of Mode 1 was approximately equal to the energy range on IN16B, thus being well described by the flat background term. Resolution measurements were obtained at 2 K on each spectrometer, the IN5 resolution has a systematic asymmetric and stepped profile intrinsic to the spectrometer.

**QENS data analysis and fitting**

All data were analysed using Mantid v6.6.0. To calculate the EISF of each of the two modes above the phase transition observed on IN16B and IN5 we modified Eq. 4 to separate the slow and fast components. The EISF of the slow component was approximated according to:

$$\text{EISF}_{\text{slow}}(Q) = \frac{I_{\text{elastic}}(Q)}{I_{\text{elastic}}(Q) + I_{\text{inelastic,slow}}(Q)}.$$

Likewise, we determined the EISF of the fast component according to:

$$\text{EISF}_{\text{fast}}(Q) = \frac{I_{\text{elastic}}(Q) + I_{\text{inelastic,slow}}(Q)}{I_{\text{elastic}}(Q) + I_{\text{inelastic,slow}}(Q) + I_{\text{inelastic,fast}}(Q)},$$

by including the intensity of the slow component in the numerator, it is equivalent to performing this measurement of the fast component on a spectrometer with energy resolution of the slow component's FWHM linewidth. This method of separating EISF components has been used successfully in previous studies of NPG[12] and NH$_4$BH$_4$ [22], where it successfully distinguished distinct rotational modes from within a single QENS signal.

Simultaneous global fitting of FWS and QENS data was performed to elucidate the behaviour of the modes above the phase transition. Four datasets were included in the fit (IFWS at 3 µeV energy transfer, and three QENS scans (320 K, 330 K and 340 K), each containing spectra for 16 different $Q$ values. Equation 1 was used directly as the model function for QENS, representing an elastic peak, two Lorentzian components and a flat background, all convoluted with the measured resolution function of the instrument. For the IFWS, Eq. 1 was modified by removing the elastic contribution and modelling a temperature dependence by choosing an Arrhenius law for the linewidths, $\Gamma_i(Q)$:

$$\Gamma_i(Q) = \Gamma_{0_i}(Q) e^{-\frac{E_{a_i}}{k_b T}}.$$

Subsequently, parameter ties were established between the different datasets to achieve the following:
- A global, $Q$-independent value for the activation energy of each process ($E_{a_1}$ and $E_{a_2}$).
- Consistency between the IFWS linewidths, calculated using the Arrhenius law above, and the fitted Lorentzian linewidths $\Gamma_i(Q)$ from the QENS spectra at different temperatures.
- Identical ratios of the amplitudes of Lorentzian contributions ($A_1(Q) / A_2(Q)$) across all spectra (QENS and IFWS) with the same $Q$.

## Results and Discussion

### Theory

QENS is used to measure inelastic cold neutron scattering from materials at low energy transfer of a few meV. Datasets appear as an elastic peak that is broadened by quasielastic scattered intensity due to energy exchange with rotational and translational diffusion processes within the sample[23]. Thus, it is possible to use QENS to track low energy dynamic modes occurring within a sample by inspecting and characterising the different quasielastic signals. The accessible timescales of these processes are primarily determined by the energy resolution of the spectrometer used to probe the sample, where better energy resolution enables the observation of processes with slower timescales (up to a few ns). Here, we utilised two spectrometers, IN16B and IN5 at Institut Laue-Langevin (ILL), which each have a distinct energy resolution and dynamic range (see Methods for instrument characteristics). This made it possible to access energy transfers from meV to µeV, corresponding to fast (ps) and slow (ns) processes, respectively.

The measurable quantity in a QENS experiment is the dynamic structure factor $S(Q,\omega)$, which provides intensity as a function of the momentum ($Q$) and energy transfer ($\omega$) of neutrons interacting with the sample. In NPG, scattering from hydrogen atoms ($^1$H) dominates the QENS signal due to their large incoherent scattering cross-section. $S(Q,\omega)$ can therefore be seen as a direct measurement of dynamic processes involving the motion of $^1$H. Based on the geometry of the NPG molecule and a previous theoretical study[24], we identify three accessible rotational modes in the PC phase. Mode 1 relates to rotation of the methyl (-CH$_3$) groups about the C-C bond axis. Mode 2 describes the similar rotation of hydroxymethyl (-CH$_2$OH) groups about the C-C bond axis. Mode 3 is a more complicated rotational reorientation of the whole NPG molecule about its centre of mass. In the OC phase, it is expected that only Mode 1 is accessible because the hydrogen bond network locks the hydroxyls, and hence the entire molecule, into a fixed orientation and position. These three rotational modes are shown schematically in Fig. 1a. A previous QENS study of NPG[12] was able to detect two of these three modes, where the third was not resolvable due to its slower timescale[24].

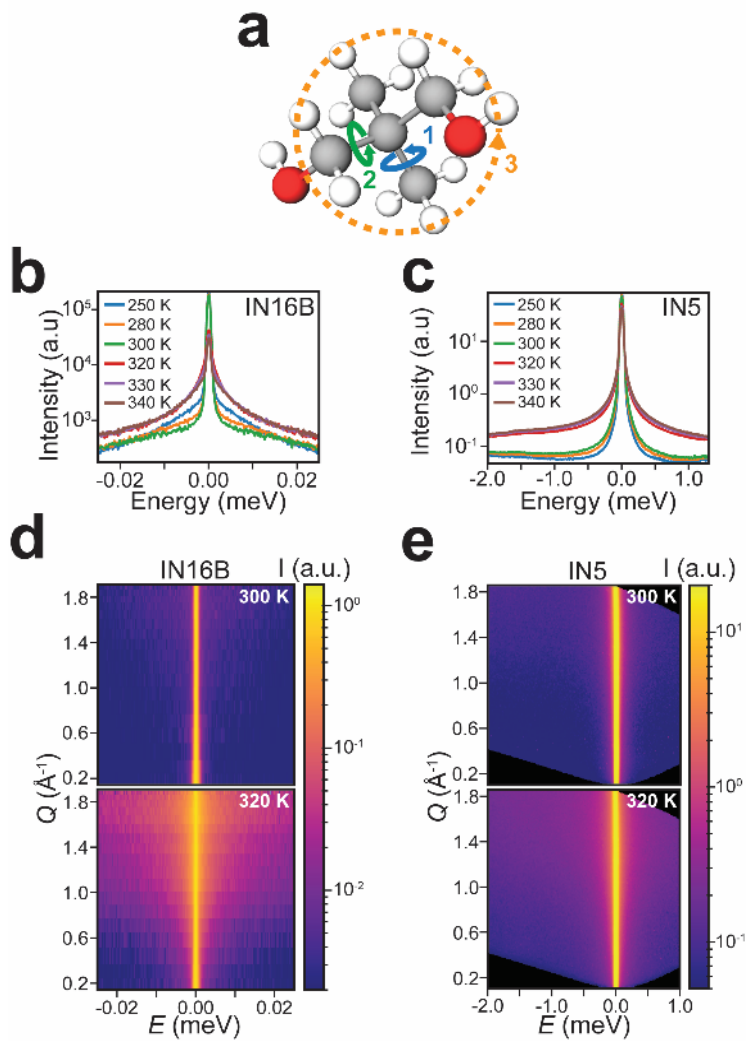

Figure 1 | **QENS data of NPG above and below its solid-solid phase transition.** (a) Geometry of $^1$H dynamical modes in NPG (b,c) QENS line profiles – ie. $S(\omega)$ integrated over $Q$ for each measured $T$ – obtained from IN16B and IN5 instruments, respectively. Note the different energy axis scales. (d,e) $S(Q,\omega)$ maps below (300 K, top) and above (320 K, bottom) the phase transition, measured on IN16B and IN5 instruments, respectively.

The measured QENS data, $S_{measured}(Q,\omega)$, obtained for NPG and shown in Fig. 1d-e derives from a convolution of the instrumental resolution function, $R(Q,\omega)$, and the theoretical incoherent scattering function, $S(Q,\omega)$, through [25]:

$$S_{\text{measured}}(Q,\omega) = R(Q,\omega) \otimes [C(Q)S(Q,\omega)] + B(Q), \tag{1}$$

where

$$S(Q,\omega) = A_0(Q)\delta(\omega) + \sum_{i=1}^{n} L_i, \tag{2}$$

and

$$L_i = A_i(Q)\frac{1}{\pi}\frac{\frac{1}{2}\Gamma_i(Q)}{\omega^2 + (\frac{1}{2}\Gamma_i(Q))^2}. \tag{3}$$

In Eq. 1, $C(Q)$ is related to the Debye-Waller factor arising from stochastic displacement of $^1$H from lattice vibrations. The effect of $C(Q)$ will be omitted in the characterisation of quasielastic processes, however determination of the mean squared displacement (MSD) associated with this stochastic motion will be treated at the end of the Results section. $B(Q)$ is a flat background term accounting for inelastic (high-energy transfer) contributions. All of the coherent and incoherent scattering arising from the sample is contained within $S(Q,\omega)$. In Eq. 2, the first term corresponds to the elastic component, modelled by a delta function, $\delta(\omega)$, with amplitude $A_0$; and the second term corresponds to the quasielastic component, modelled by a sum of Lorentzians (given by Eq. 3), with amplitudes $A_i$, each corresponding to a distinct dynamical process with characteristic linewidth ($\Gamma_i$). The associated timescale, $\tau_i$, of each of these processes can be found through the relation $\tau_i = \frac{2\hbar}{\Gamma_i}$. It should be noted that if $\Gamma_i$ is much smaller than the width of the instrumental energy resolution, then the corresponding mode will contribute to the recorded elastic intensity, as if the associated rotation was stationary; whilst if $\Gamma_i$ is substantially larger, then the corresponding mode will contribute to the measured background.

For NPG in its OC phase, we expect the measured QENS signal to resolve quasielastic contributions arising from Mode 1 only, so that Eq. 2 has only one Lorentzian term, while Modes 2 and 3 are also present in the in the PC phase, yielding three Lorentzian terms. As outlined below, Mode 1, the methyl group rotation, is typically simplified to model $^1$H jumps between three equivalent sites equally spaced on a circle.[25] Similarly, we will model mode 2 as continuous rotation on a circle to track the hydroxyl $^1$H motion. Mode 3, relating to a full tumbling rotational motion of the NPG molecule, is modelled as continuous, isotropic rotation on a sphere. In each of these models, the $^1$H atoms hop instantaneously between equilibrium sites with a mean time between jumps given by $\tau_i$.

The amplitude of the elastic signal given by $A_0(Q)$ is known as the elastic incoherent structure factor (EISF), which is traditionally calculated as the measured proportion of elastic scattering within the total scattered intensity:

$$\text{EISF}_{\text{observed}}(Q) = \frac{I_{\text{elastic}}(Q)}{I_{\text{elastic}}(Q) + I_{\text{inelastic}}(Q)}. \tag{4}$$

When considering contributions from a single quasielastic component, the EISF can be calculated simply using Eq. 4. However, if more modes are resolved, then a distinct EISF for each inelastic component may be determined[12,22] (see Methods for details). The spatial geometry of the observed dynamical process leads to a characteristic functional form of EISF$_{\text{observed}}(Q)$, allowing it to be used to identify dynamics occurring within the sample. Finally, we can modify Eq. 4 to account for the possibility that only a fraction of scatterers participate in the observed process with a given geometry, modelled with EISF$_{\text{model}}(Q)$:

$$\text{EISF}_{\text{observed}}(Q) = (1 - f) + f\,\text{EISF}_{\text{model}}(Q), \tag{5}$$

where $f$ is the fraction of scatterers mobile on the timescale of the spectrometer [26–28]. In Eq. 4 & 5 the nature of the $Q$ dependence is directly related to the geometry of the motion of $^1$H. Thus, distinct geometric models (EISF$_{\text{model}}(Q)$) can be used to fit the calculated EISF$_{\text{observed}}(Q)$, to determine the type of motion that produces the measured QENS signal.

## Results & Discussion

Figure 1b,c shows the QENS data obtained from NPG upon heating through its S-S phase transition ($T_0$ = 314 K), from 250 K to 340 K. The data are integrated across all measured $Q$ for each spectrometer, with full width at half maximum (FWHM) energy resolutions of 0.75 µeV for IN16B (Fig. 1b) and 60 µeV for IN5 (Fig. 1c). It is evident that both instruments detect a significant broadening of the elastic peak through the S-S phase transition, and hence record an increase in the quasielastic signal that is made more apparent in the full $S(Q,\omega)$ data given in Fig. 1d,e. This increase in signal is due to additional dynamical modes liberated in the PC phase. The IN5 data in Fig. 1c show the phase transition as a jump in inelastic intensity between 300 K and 320 K. Furthermore, these data visually show only a weak variation of broadening with $Q$, which suggests localised dynamics, as discussed later. After the phase transition occurs, there is a large increase in quasielastic signal across all measured energies. The data from IN16B in Fig. 1b provide more detail of the broadening at low energy transfers. Below the phase transition, we observe a decrease in quasielastic intensity within ±0.01 meV of the elastic peak from 250 K to 300 K, suggesting that the frequency of this process increases with increasing temperature. Again, after the phase transition we observe an increase in quasielastic signal, with a weak dependence on $Q$ now visible across the narrow energy range accessed by IN16B, shown in Fig. 1d. The full energy range QENS data measured on IN5 can be found in Supporting Information Fig. S2.

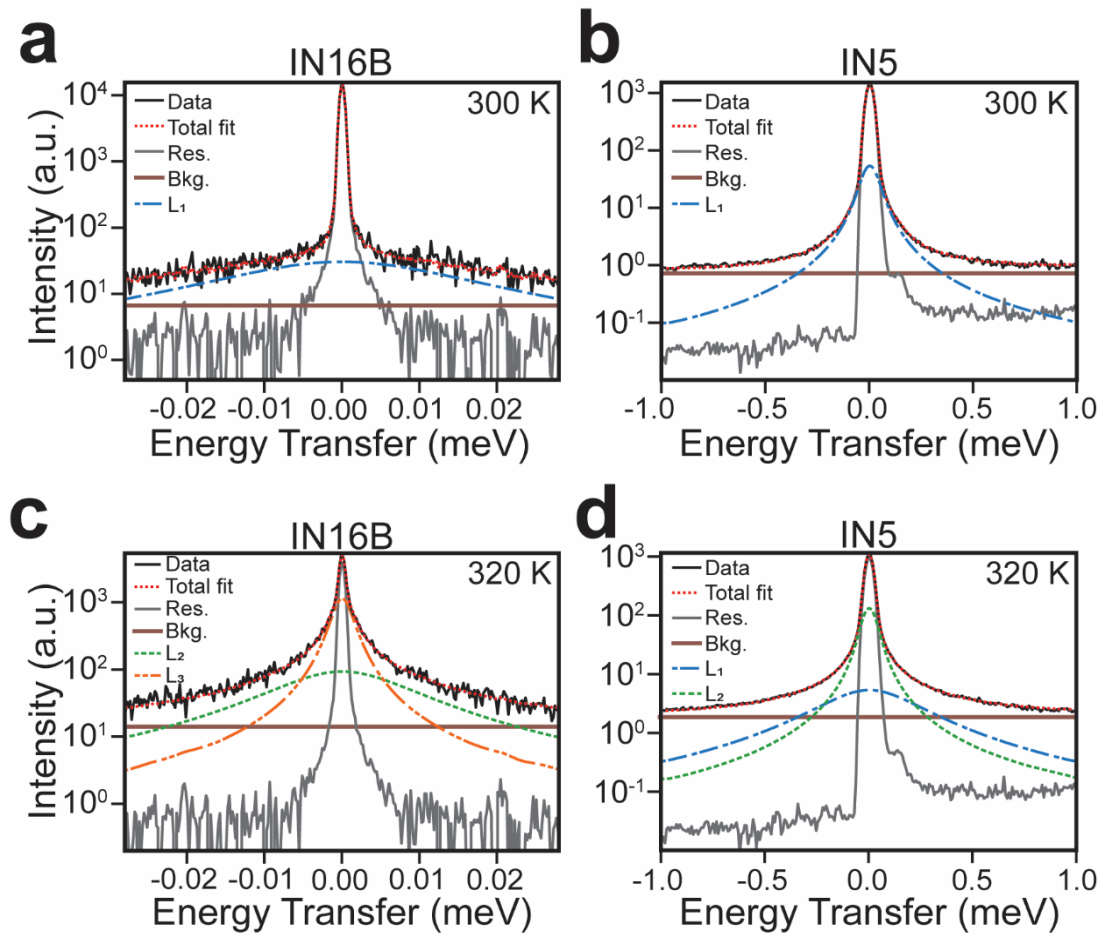

**Figure 2 | Representative QENS data fits for Q = 0.83 Å$^{-1}$.** (**a,b**) Example fits for IN16B and IN5 QENS data below the phase transition (300 K). Here, only a single Lorenztian is included in the model. (**c,d**) Example fits above the phase transition (320 K). In this case, two Lorenztians are required to fit the data. Sequential fitting across all Q was performed on each dataset. The colour and subscript of the Lorenztian fits is determined by the identification of the mode using EISF fitting in Fig. 4. The asymmetric profile of the resolution function for IN5 in (**b,d**) arises from the spectrometer construction and is systematic across all datasets.

To determine the dynamic processes contributing to the quasielastic signal at each measured $T$, the full experimental dataset was fitted using Eqs. 1 and 2. Fitting was performed sequentially to each $Q$ range within the dataset, and example fits corresponding to $Q = 0.83$ Å$^{-1}$, which is representative of the other $Q$ values, are presented in Fig. 2. In each case, it is apparent that there is a good fit to the data using the model. Note that the colours of the Lorentzian components are chosen to specify distinct modes, as determined through $Q$-dependence of the EISF (Eq. 5), which will be presented below. The resolution function, $R(\omega)$, for each value of $Q$ has been measured directly by recording a data set at 2 K. This is the reason for the grey traces in Fig. 2 appearing noisy; and also, the reason for the asymmetric and stepped profile for the IN5 resolution function, which reveals some experimental artefacts at low signal levels that do not affect subsequent analysis.

We expect the mode identified at 300 K, ie. below $T_0$, to be the methyl group rotation. In the data from IN16B (Fig. 2a) the fitted $L_1$ term is very broad, approaching the limits of the instrumental dynamic range, and so relates to a relatively high-frequency mode. In contrast, the same feature is detected on IN5 more clearly (see Fig. 2b). Fig. 2c,d shows similar QENS data and fits above $T_0$, at 320 K. For IN16B, in Fig. 2c, the data are fitted well with two Lorentzian terms, one ($L_2$) with a linewidth of the same order of magnitude as $L_1$ in Fig. 2a, whilst the other mode ($L_3$) is significantly narrower. For IN5, in Fig. 2d, the data are also fitted with two Lorentzians, but the linewidth of the first ($L_1$) is orders of magnitude larger than that of either mode measured on IN16B at 320 K. Although the linewidth of the second component ($L_2$) is comparable to that of the single mode ($L_1$) observed at 300 K on IN5 in Fig. 2b, it has been labelled as a distinct mode because the EISF data presented shortly shows distinct behaviour (see Fig. 4). These observations suggest that by considering results from both spectrometers, we can observe at least three unique dynamic processes above $T_0$.

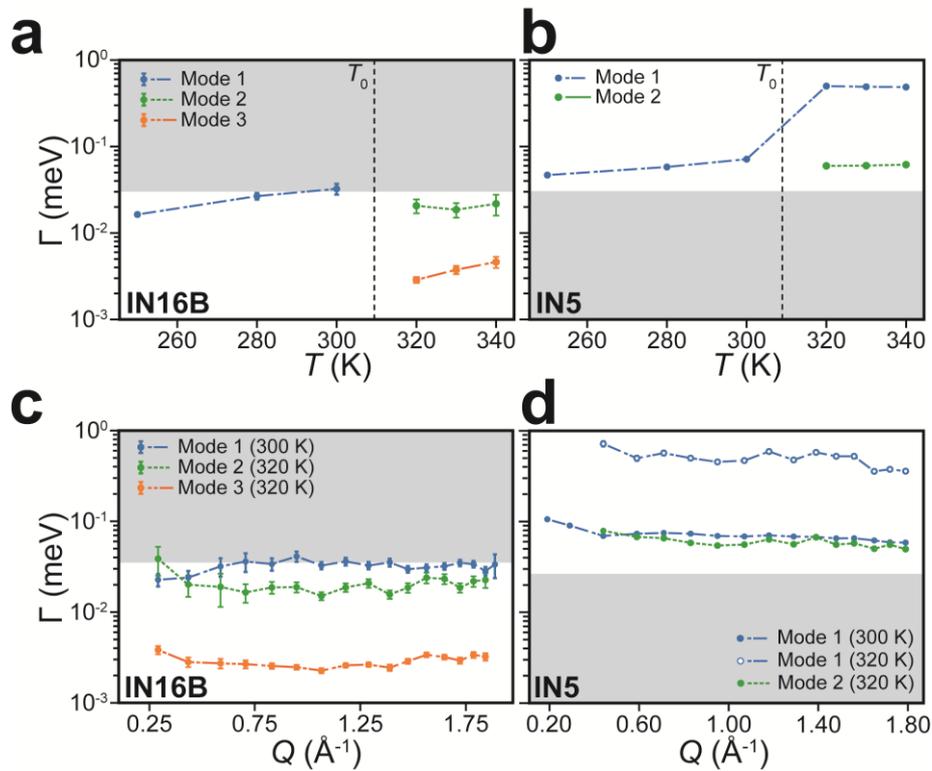

**Figure 3 | $T$- and $Q$-dependence of mode frequencies.** (**a,b**) $\Gamma(T)$ for each of the observed modes, from the Lorentzian fits across the two instruments. The black dashed line indicates the equilibrium phase transition temperature. (**c,d**) $\Gamma(Q)$ for each of the observed modes at temperatures just below (300 K) and above (320 K) the phase transition. Line colour indicates the mode identity as found from fitting EISF (Fig. 4). Grey regions represent the approximate timescales outside of the dynamic range of the spectrometer.

The $T$- and $Q$-dependences of the three modes are explored in more detail in Fig. 3. Fig. 3a,b plots $\Gamma_i(T)$ of each mode across the two spectrometers, where the grey regions represent process timescales lying outside of the dynamic range of the spectrometer. Below $T_0$, we expect only the methyl group rotation to be active, and the only detectable Lorentzian contribution has a similar linewidth, $\Gamma_1 \approx 50$ µeV and $\tau_1 \approx 26$ ps, measured across both spectrometers. Above $T_0$, we observe two additional modes: Mode 2, with $\Gamma_2 \approx 20$ µeV and $\tau_2 \approx 66$ ps, is resolved on both spectrometers whilst Mode 3, with $\Gamma_3 \approx 3$ µeV and $\tau_3 \approx 440$ ps is observed only on IN16B. The original mode, retaining $\Gamma_1 \approx 500$ µeV, remains apparent in the IN5 dataset but relates to a motion that has become too fast to be accessible to IN16B. This step-change in methyl rotation frequency across the phase transition has been observed previously[12], and is likely due to different activation energies for the rotation in the PC phase compared to the OC phase because of the commensurate lattice expansion. Conversely, Mode 3 is not detected on IN5, as the linewidth is too small, and the timescale too slow, for the motion to be resolved. Figure 3c,d then plots $\Gamma(Q)$, at 300 K and 320 K for IN16B and IN5, respectively. The trends are independent of $Q$, demonstrating that the modes are spatially localised and unlikely, for example, to represent diffusion processes[29].

To identify the rotational processes that correspond to the modes shown in Fig. 3, we determined the EISF associated with each Lorentzian. The calculated EISF data are then fitted using different geometric rotational models, using their goodness of fit and molecular geometry considerations to assign each of the observed modes. We start by presenting the rotational models, with the assumption that the quasielastic signals are arising solely from incoherent $^1$H scattering due to jump diffusion between energetically equivalent sites with distinct $\tau_i$. The methyl rotation (Mode 1) is well described by the 120° three hop (3-Hop) model, which takes the form[25]:

$$\text{EISF}_{3-\text{Hop}}(Q) = (1-f) + f\frac{1}{3}\left[1 + \frac{2\sin(\sqrt{3}Qr)}{\sqrt{3}Qr}\right], \tag{6}$$

where $r$ is the rotational radius of $^1$H motion associated with this mode. We identify Mode 2 with hydroxymethyl rotation, which should have a similar functional form to that of Mode 1, modelling the $^1$H scatterers as hopping between equivalent sites around a circle. However, recent molecular dynamics simulations[30] have shown that because the hydroxymethyl rotation lacks three-fold symmetry, it is expected that its motion is more complex (Supporting Information Fig. S4 and S6), with more than three preferred orientations. We approximate this motion as hopping between six equivalent sites on a circle of radius $r$, as it has been shown that this is a good approximation for six or more hopping sites within the $Q$-range observed here[25]. The EISF for this continuous hop (C-Hop) model is taken as[25]:

$$\text{EISF}_{\text{C-Hop}}(Q) = (1-f) + f\frac{1}{6}\left[1 + 2\frac{\sin(Qr)}{Qr} + 2\frac{\sin(\sqrt{3}Qr)}{\sqrt{3}Qr} + \frac{\sin(2Qr)}{2Qr}\right]. \tag{7}$$

For ease, Finally, the EISF of Mode 3 is fitted assuming reorientation of the whole molecule, using a model that describes $^1$H atoms continuously diffusing on a sphere with radius $r$[25]:

$$\text{EISF}_{\text{reorientation}}(Q) = (1-f) + f\left[\frac{\sin(Qr)}{Qr}\right]^2. \tag{8}$$

Figure 4 presents the EISF$(Q)$ fits using the above models for the observed rotational processes across both spectrometers. The distinct functional forms and fitted radii for EISF$(Q)$ justify their identification with three distinct processes, each corresponding to a rotational mode with a unique geometry. First inspecting the data from IN16B in Fig. 4a, Mode 1 (blue circles) appears to be described well by the 3-Hop model defined in Eq. 6. By fixing $r$ = 0.92 Å as calculated from NPG crystallography data[31], the best fit yields $f$ = 0.42. Note that an upper bound of $f$ = 0.5 was imposed for this fit since half of the scattering hydrogens are stationary on the timescale of IN16B in the OC phase, and thus only contribute to elastic intensity. Next, the trend of Mode 2 (green circles) is described well by the C-Hop model defined in Eq. 7 using $r$ = 1.40 ± 0.1 Å and $f$ = 0.58, as determined by the best fit. This is slightly larger than expected from the geometry of the molecule shown in Fig. 4e. However, applying this model to the motion of the hydroxymethyl hydrogens is a substantial simplification because the hydroxymethyl group contains two distinct $^1$H types. Specifically, the hydroxyl $^1$H has an additional rotational degree of freedom about the C-O bond, indicated by the red arrow in Fig. 4e, which is not captured by the model. While the model provides a good fit to our data, and is consistent with expectations based on previous theory, further refinements to better capture hydroxymethyl rotation are expected. Finally, the trend of Mode 3 (orange circles) is described well by the reorientation model defined in Eq. 8 and yields a best fit rotational radius of $r = 2.04 \pm 0.1$ Å and $f = 0.90$. This radius agrees well with the average hydrogen distance from the molecular centre of mass if a full isotropic tumbling motion is considered, as indicated in Fig. 4f. Our assignment of Mode 2 ($\tau_2 \approx 66$ ps) as hydroxymethyl rotations and the slower Mode 3 ($\tau_3 \approx 440$ ps) as molecular reorientations is consistent with recent molecular dynamics simulations on NPG [30]. Here, the authors resolve transitions between distinct molecular orientations but not rotations of the hydroxymethyl group, as these are claimed to be occurring on a timescale shorter than the time interval used for configurational sampling. See Supporting Information for calculations of the NPG geometry and justification for the radii used in the schematics of Fig. 4d-f.

Moving on to the data collected for higher energy transfers, Fig. 4b shows the corresponding data obtained from IN5. Immediately apparent is the insensitivity of IN5 to Mode 3, which is expected since $\Gamma_3$ is smaller than the energy resolution of IN5. The fit for Mode 1, again fixing $r = 0.92$ Å, provides $f = 0.44$, which is very close to the value obtained on IN16B. For Mode 2, we again fit these data using the C-Hop model to represent the hydroxymethyl rotation yielding $r = 1.59 \pm 0.1$ Å. This fit agrees well with the Mode 2 fit from IN16B data in Fig. 4a. We also note that the data presented in Fig. 4b agrees well with previously obtained QENS data for NPG using a spectrometer with comparable energy resolution to IN5[12]. That study utilised the neutron spectrometer AMATERAS with an energy resolution of 50 µeV, but the authors instead assigned these data to methyl rotation and molecular reorientation, respectively.

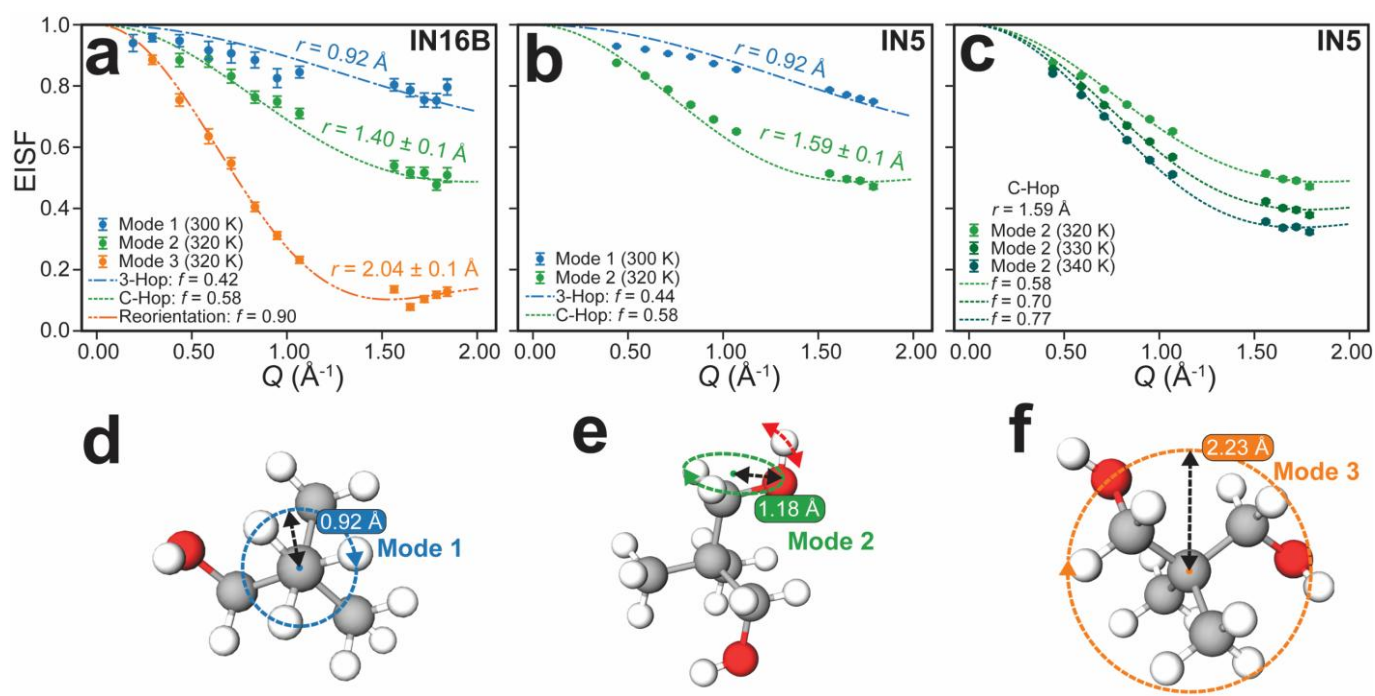

**Figure 4 | Fitting of EISF(Q) and identification each of the observed modes.** (**a,b**) Summary of EISF fitting for T = 300 K, 320 K to data collected from (**a**) IN16B and (**b**) IN5. Here, three modes are identified and fit to incoherent scattering models as defined in the text. (**c**) T-dependence of the hydroxymethyl mode fraction fitted to IN5 data. (**d-f**) Schematics of the NPG molecule, indicating the geometric origin for the radii of rotation obtained from fits in (**a,b**). Data between $1.1 < Q < 1.5$ Å$^{-1}$ was masked from fits due to the presence of Bragg peaks in this Q-range.

The Mode 2 fractional fits for both IN16B and IN5 seem to suggest that at 320 K we detect 116% of the hydroxymethyl rotational motion, with an expected upper limit of $f = 0.5$. This further indicates that the model is simplified for this mode. To determine how $f$ varies with temperature on IN5 we fix $r = 1.59$ Å and refine $f$, the result of which can be seen in Fig. 4c. Here, we observe a very clear trend of increasing $f$ from $T = 320$ K to 340 K. A possible explanation is that at higher temperatures, quasielastic intensity from the broadening
9

reorientation motion is appearing in this model. The remaining EISF plots for other measured temperatures can be found in Supporting Information Fig. S3.

The EISF analysis presented in Fig. 4 highlights the importance of the fractional fitting parameter, $f$, to provide meaningful fits to all the observed modes. The inclusion of this parameter has been used extensively to study other material systems [26–28,32], where it is often interpreted as representing a fraction of 'bound' or otherwise 'non-contributing' scatterers. Here, we would like to offer possible physical interpretations of this parameter in the fitting of the isotropic molecular reorientation (Mode 3) in NPG using the high-resolution IN16B data shown in Fig. 5a. The motion has been previously described as a 'complex reorientation mode' with a non-trivial energy landscape, consisting of rotation about an axis that is also slowly rotating[12,24]. The characteristic timescale of $\Gamma_3$ is ~650 ps at 320 K, much faster than the time-resolution limit of IN16B (~ 4 ns). Nevertheless, Fig. 5a suggests that the fraction of observed scatterers using IN16B increases with $T$. By extrapolating this trend, in Fig. 5b we assess how the scatterers are contributing to the mode at $T_0$ and beyond. It is a subtle effect but the fitted fraction, $f$, increases from 0.89 to 1.00 between 314 K and 420 K. The latter temperature is close to the melting temperature of NPG (402 K), where the molecules obtain additional translational degrees of freedom. This observation indicates that 1) not all the NPG molecules are freely rotating immediately after the phase transition and/or 2) not all orientations are equally accessible to the molecules at the phase transition. The first possibility could be attributed to structural inhomogeneities in the material, for example, grain boundaries recently characterised using polarised light[15] and electron[16] microscopy, that result in different local environments and geometric hindrance of the molecules. Another explanation for the first possibility is that at lower temperatures, transient hydrogen bonds lock a small fraction of molecules in place for some time. The second possibility agrees with previous theoretical work that showed that the activation energy barrier for molecular reorientation is not isotropic[24].

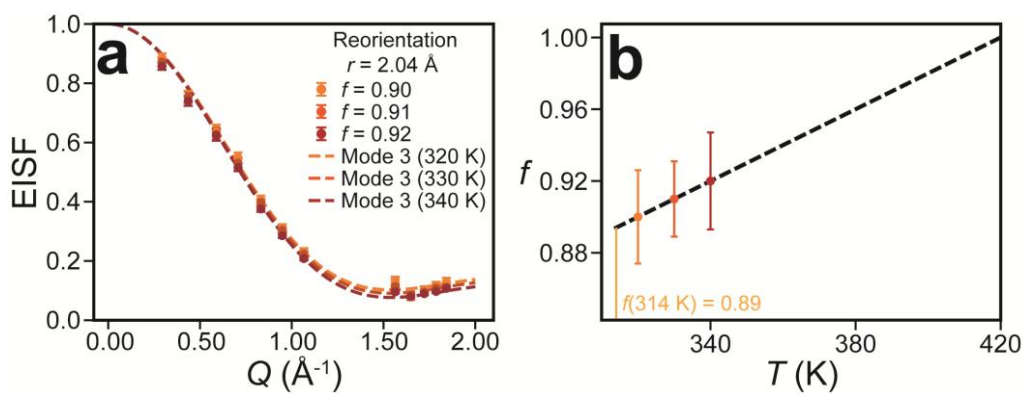

**Figure 5 | Reorientation mode fraction as a function of temperature.** (**a**) EISF($Q$) plots of the reorientation mode for temperatures above the phase transition. (**b**) Reorientation mode fraction, $f$, as a function of $T$, line of best fit allows for the estimation of $f$ = 0.89 at the phase transition, $T_0$ = 314 K. This mode is fully liberated at 420 K, as determined by the best fit at $f$ = 1.0.

Our observation that the rotational motion of NPG molecules is not fully liberated at $T_0$ has implications for efficient operation of NPG as a BC. Operating very close to $T_0$ at around 320 K (see Supporting Information Fig. S1 and Ref. [33]) will not maximise the entropy changes associated with the phase transition. It is also important in estimating configurational entropy changes around the phase transition. Such calculations allow an estimation of the BC performance of PCs but can both greatly overestimate[24] or underestimate[34] the entropy changes obtained from calorimetry measurements. A more gradual liberation of molecular reorientations may explain why the calculated configurational entropy change, $\Delta S = R\ln(180) = 415$ J K$^{-1}$ kg$^{-1}$, is overestimated by around 6%. Taking instead the configurational entropy change to be $\Delta S = R\ln(180*0.89)$, essentially reducing the number of accessible microstates at $T_0$, yields a value of 405 J K$^{-1}$ kg$^{-1}$, closer to the experimentally determined value of ~390 J K$^{-1}$ kg$^{-1}$ [12,33,35–37], and consistent with the value of 400 J K$^{-1}$ kg$^{-1}$ recently obtained from molecular dynamics simulations of NPG[30]. Of course, this estimation only considers the configurational entropy of the molecules and does not include other minor contributions such as lattice distortions. Regardless of the true physical interpretation of *f*, this study indicates that configurational entropy change calculations should not be solely relied upon for estimating the BC performance in candidate materials.

Finally, we turn our attention to the fixed window scans (FWSs) using the IN16B instrument. This unique technique allows the QENS signal to be monitored at specific energy transfers during thermal processing of a sample. Figure 6a shows the elastic FWS (EFWS) and Fig. 6b shows the inelastic FWS (IFWS) for NPG on heating from 280 K to 345 K (red data) and cooling from 345 K to 10 K (black data) for $Q = 0.83$ Å$^{-1}$. The elastic signal corresponds to the scattered intensity about zero energy transfer within instrumental energy resolution (± 0.75 µeV). Similarly, the IFWS was collected with the Doppler monochromator set up to collect inelastic signal corresponding to an energy transfer of ±3 μeV. The two data sets complement one another. A loss of EFWS intensity is accompanied by an increase in inelastic scattering and so the general trend in Fig. 6a shows gradual thermal activation of vibrational and rotational modes in NPG. However, because the IFWS records only a narrow window, it does not simply show the inverse of the EFWS, and the energy window can be selected to accentuate differences in the QENS peak broadening. Here, the window chosen focuses on the clear differences in the peak shoulder observed between 250 K and 320 K in Fig. 1b, where the intensity appears to first drop below $T_0$ and then suddenly increase above $T_0$.

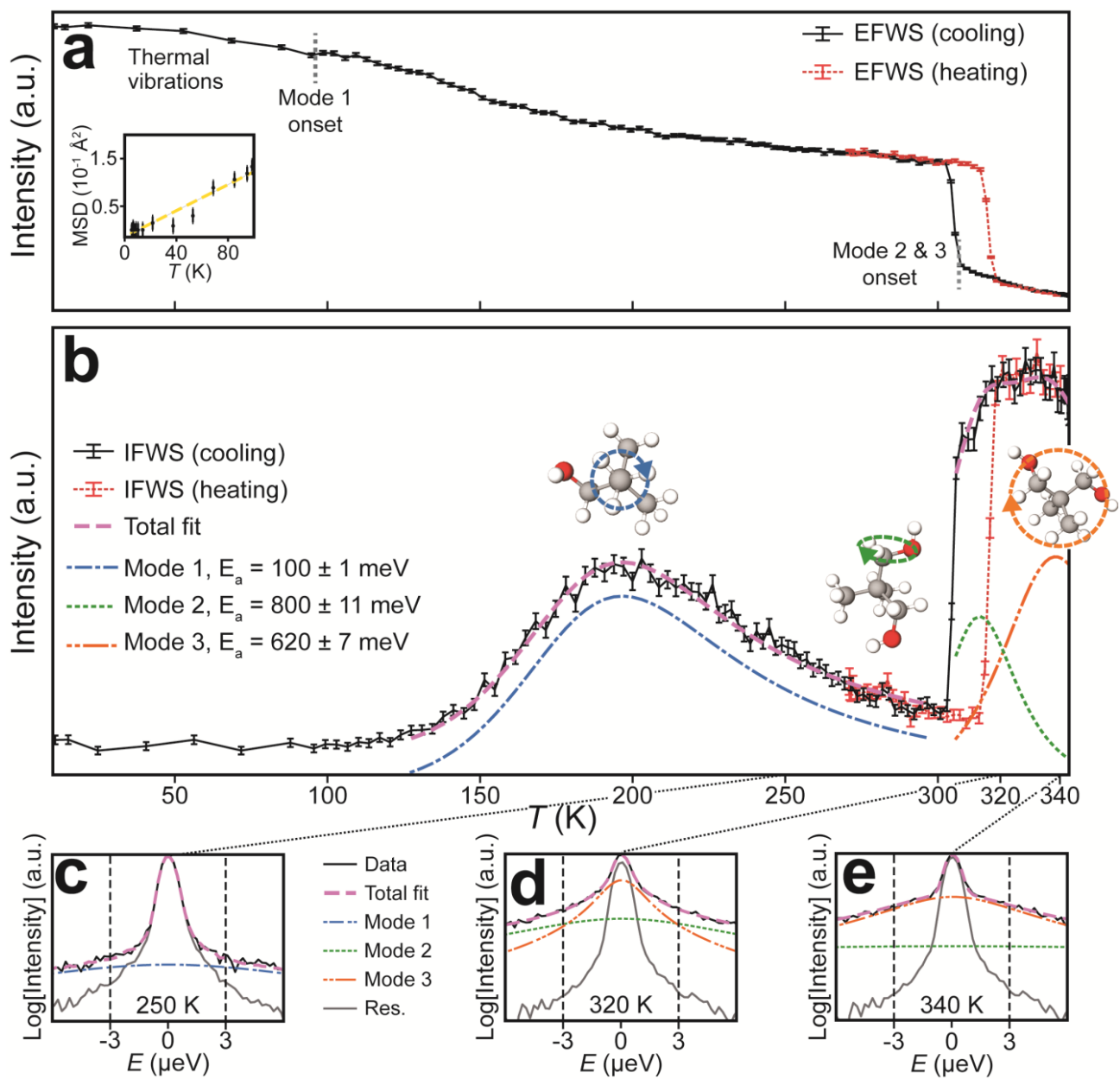

**Figure 6 | Combined FWS and QENS global fitting analysis.** (**a**) EFWS (0.0 µeV) for NPG on heating and cooling. Inset MSD(*T*) obtained from fitting of EFWS for *T* < 100 K. (**b**) IFWS (± 3.0 µeV) for NPG on heating and cooling. Fitting of IFWS peaks on cooling at 200 K, 310 K and 340 K correspond to methyl, hydroxymethyl and molecule reorientations, respectively, as shown by the inset schematics. Global fits including QENS scans shown in (c-e) provide measured activation energies of 100 meV, 800 meV and 620 meV respectively. All data shown is for $Q = 0.83$ Å$^{-1}$, representative of other *Q* values. See Methods for fitting details. The width of the energy window for both EFWS and IFWS is ± 0.75 µeV.

Between 10 K and 100 K, there is little IFWS signal in Fig. 6b, since any rotational motion occurs on a timescale too slow to be observed. The simultaneous, slow reduction in EFWS (Fig. 6a) arises from thermal vibrations and librations associated with the $^1$H atoms. This temperature region was fitted across all $Q$ to obtain the Debye-Waller factor mean squared displacement (MSD) associated with the low-$T$ thermal motion, shown in the inset of Fig. 6a. The linear best-fit to this data provides the temperature-dependence of this motion as ~1.4 × 10$^{-3}$ Å$^2$ K$^{-1}$, which tentatively equates to a MSD at 300 K of 0.4 Å$^2$. At 100 K, the inelastic signal in Fig. 6b increases as the elastic signal in Fig. 6a starts to decrease, both due to the appearance of the methyl group rotation (Mode 1). This gradual loss of elastic intensity continues across the 100 K – 300 K range, yielding a distinct peak in the IFWS signal in Fig. 6b. This peak can be attributed to the $L_1$ term 'moving through' the fixed energy window as the associated mode decreases in frequency with decreasing temperature. By fitting the thermal variation of the $L_1$ term (blue line) and including a flat background (not shown for visual clarity), as described previously and assuming an Arrhenius behaviour[38], we obtain an activation energy for Mode 1 of 100 ± 1 meV. Fig. 6c shows the QENS scan obtained at 250 K and demonstrates how the IFWS signal is obtained at the dashed lines located at ± 3 μeV.

Both FWS plots in Fig. 6a,b show clear hysteresis in the modes detected by inelastic neutron scattering. At ~314 K on heating there is a large discontinuous decrease in elastic signal in Fig. 6a, consistent with the phase transition temperatures determined by calorimetry (Supporting Information Fig. S1), corresponding to the increase in inelastic scattering through the S-S phase transition. A similar abrupt increase in elastic signal occurs instead at ~304 K upon cooling in Fig. 6a, corresponding to the S-S phase transition driven by the eventual formation of the hydrogen bond network. The step-change in inelastic signal in Fig. 6b highlights the fact that on heating (cooling), Modes 2 and 3 disappear (appear) on the breaking (formation) of the hydrogen bond network. In NPG, the appearance of these modes implies the loss of the hydrogen bond network that is mediated by the hydroxyls; hydroxymethyl rotation and full molecular reorientation can only occur when the hydrogen bond network is lost.

On cooling from 343 K to 304 K, the IFWS intensity passes through a maximum at ~320 K, which we attribute to the $L_2$, $L_3$ terms narrowing, demonstrated by the QENS data in Fig. 6d,e. This reduction in the frequency of Modes 2 and 3 occurs upon cooling until a temperature of ~304 K, where the $L_2$, $L_3$ terms vanish, the tumbling motion and hydroxymethyl rotation both stop, and the molecules are locked back into a hydrogen-bonded lattice. With an energy offset of 3 μeV, the IFWS data above 304 K in Fig. 6b contains signal from both $L_2$ and $L_3$ terms. This can be confirmed with the QENS scans in Fig. 6d,e at 320 K and 340 K, respectively, where $L_2$ (green line) and $L_3$ (orange line) intercept the dashed lines at ± 3 μeV. Using this observation, we combine the IFWS data with the QENS scans at 320 K, 330 K and 340 K in a global fit to characterise the behaviour above the S-S phase transition across the full $Q$-range (see Methods for fitting details). Additional IFWS and QENS fits for low- and high-$Q$ values can be seen in Supporting Information Fig. S7. This fitting procedure yields activation energies of 800 ± 11 meV for Mode 2 and 620 ± 7 meV for Mode 3. The relative magnitudes of these values match those calculated by others from simulations[24] and suggest that, surprisingly, the hydroxymethyl rotation has a higher activation energy than the molecular reorientation in the PC phase.

Our global fitting procedure allows us to extract the individual behaviours of the hydroxymethyl rotation and molecular reorientation in the hysteresis region (304 K to 314 K). Specifically, on cooling, the linewidth of the $L_3$ term (orange line) narrows to below the 3 μeV energy window offset, approaching the energy resolution of the spectrometer. This indicates that the molecular reorientations continue to lower temperatures than during the heating cycle, but cease as the hydrogen bond network is formed. Conversely, the linewidth of the $L_2$ term (green line) contributes almost all the IFWS signal below ~310 K, indicating that hydroxymethyl rotations are more active despite the significant slowing of whole molecule rotations.

Our results demonstrate that IFWS and EFWS techniques provide a direct experimental method for probing orientational degrees of freedom in molecular crystals, closely related to the presence of hydrogen bond networks. Similar recent studies are often carried out using less direct techniques such as Raman spectroscopy [24,39–41], calorimetry[13,33,42,43] or QENS analysis at a small number of discrete temperatures[12,28,34,44]. Here, we are able to measure the material behaviour continuously through thermal cycles, giving direct observation of molecular dynamics within hysteretic regions of a phase transition whilst simultaneously extracting key physical parameters such as activation energies.

For the case of neopentyl glycol derivatives and related PCs, it is clear that supercooling and hysteresis effects are dependent on the establishment of a hydrogen bond network. Although hysteresis effects can be expected within a first order order-disorder phase transition, there is scope to reduce that hysteresis and the associated stochastic supercooling effects and thereby improve the viability of NPG as a commercial refrigerant. In the case of NPG, the entropic and enthalpic changes exploited by a refrigeration cycle derive from the onset of full rotational disorder of the NPG molecules. There is, however, no *a priori* reason to assume that full molecular rotation should be commensurate with activation of the hydroxymethyl rotation since the energy landscapes for these two excitations are different. Nevertheless, we demonstrate here that the latent heat associated with the phase transition is unlocked by the activation of the hydroxymethyl rotational mode. Once active, full molecular reorientation is facile because the hydrogen bond network is simultaneously disrupted and the thermodynamic barriers to molecular reorientation are smaller than those that constrain the hydroxymethyl group. Evidence for this is already hinted at through (a) the drastic increase in transition temperature with increased number of hydroxymethyl groups in the wider group of hydroxylated neopentane PCs and (b) theoretical modelling of NPG showing that molecular reorientations are of lower energy than hydroxymethyl rotations[24]. Our results suggest a route for intrinsic control of hysteresis through modification of the energy landscape for hydroxymethyl rotation. For example, short-range disruption of the hydrogen bonds, through blending plastic crystals with different numbers of hydroxymethyl, amino, or other polarised functional groups capable of participating in hydrogen bonding. We propose the introduction of a fraction of molecules into the hydrogen bonded network that have lower barriers to rotation. These could provide the nucleation events required to suppress hysteric supercooling effects that are the only practical hurdle in the development of barocaloric refrigeration devices.

## Conclusions

By considering data from two complementary neutron spectrometers (IN5 and IN16B), we have directly tracked the molecular dynamics underpinning barocaloric performance in the benchmark barocaloric material, neopentyl glycol. Utilising the increased energy range of the complementary spectrometers, we experimentally resolve for the first time the three rotational modes of NPG: methyl, hydroxymethyl and molecular reorientation. A clear link is observed between the disruption of the hydrogen bond network and the solid-solid phase transition, in good agreement with previous studies on NPG. The supercooling effect, where the 'freezing' transition occurs at a lower temperature than 'melting', arises because residual hydroxymethyl rotations suppress hydrogen bonding. Thus, modification of the energy landscape associated with hydroxymethyl rotations may provide opportunities for engineering reduced thermal hysteresis in future materials. Our analysis permits an estimate of the fraction of molecules actively reorientating as a function of temperature, providing insight into previous discrepancies between experimental and theoretical configurational entropy change calculations. Global fitting of combined high-resolution inelastic fixed window scans and QENS measurements enables detailed insight into the temperature dependence of the detected modes from 10 K to well above the solid-solid phase transition. This allows for the extraction of mode activation energies in good agreement with previous theoretical work. Within the field of barocalorics, these results provide a deeper understanding of the detrimental hysteresis of the first-order barocaloric phase transition in neopentyl glycol. This insight can be exploited to tailor the microscopic mechanisms of the phase transition in barocaloric molecular crystals towards reducing thermal hysteresis during heating or cooling cycles. More generally, our study demonstrates the power of using a combined inelastic fixed window and quasielastic neutron scattering approach to build a more complete understanding of the molecular dynamics that drive the phase transitions and thermal hysteresis critical to the functionality of many first-order phase transition materials.

## Supporting Information

NPG calorimetry data; full energy range IN5 data; additional EISF data; NPG molecular geometry; additional IFWS fits.

## Author Contributions

Formal Analysis: F.R.B., M.A., D.B.; Investigation: F.R.B., M.A., D.B., C.I.; Conceptualisation: D.B., F.R.B., D.M.; Supervision: D.B., D.M.; Writing – original draft: F.R.B., D.B., D.M.; Writing – review and editing: all authors.

## Acknowledgements

We are grateful for insightful discussions with Lesley Cohen (Imperial College London) throughout this work. We acknowledge beam time awarded by the ILL neutron source (proposal 7-02-221, data doi: 10.5291/ILL-DATA.7-02-221). The authors thank J. Ollivier for experimental support. This work was financially supported by an EPSRC grant (EP/V042262/1). C.I. was funded by a DTP studentship 2748957 from EP/W524359/1. For open access, the authors have applied a Creative Commons Attribution (CC BY) licence to any Author Accepted Manuscript version arising from this submission.

## References

(1) Terry, N.; Galvin, R. How Do Heat Demand and Energy Consumption Change When Households Transition from Gas Boilers to Heat Pumps in the UK. *Energy Build* **2023**, *292*, 113183. https://doi.org/10.1016/j.enbuild.2023.113183.

(2) Ougazzou, M.; El Maakoul, A.; Khay, I.; Degiovanni, A.; Bakhouya, M. Techno-Economic and Environmental Analysis of a Ground Source Heat Pump for Heating and Cooling in Moroccan Climate Regions. *Energy Convers Manag* **2024**, *304*, 118250. https://doi.org/10.1016/j.enconman.2024.118250.

(3) Abas, N.; Kalair, A. R.; Khan, N.; Haider, A.; Saleem, Z.; Saleem, M. S. Natural and Synthetic Refrigerants, Global Warming: A Review. *Renew. Sustain. Energy Rev* **2018**, *90*, 557–569. https://doi.org/10.1016/j.rser.2018.03.099.

(4) National Audit Office. Decarbonising Home Heating. *Department for Energy Security & Net Zero* **2024**, 1–51.

(5) Cirillo, L.; Greco, A.; Masselli, C. Cooling through Barocaloric Effect: A Review of the State of the Art up to 2022. *TSEP* **2022**, *33*, 101380. https://doi.org/10.1016/j.tsep.2022.101380.

(6) Moya, X.; Kar-Narayan, S.; Mathur, N. D. Caloric Materials near Ferroic Phase Transitions. *Nat Mater* **2014**, *13* (5), 439–450. https://doi.org/10.1038/nmat3951.

(7) Kitanovski, A.; Plaznik, U.; Tomc, U.; Poredoš, A. Present and Future Caloric Refrigeration and Heat-Pump Technologies. *IJR* **2015**, *57*, 288–298. https://doi.org/https://doi.org/10.1016/j.ijrefrig.2015.06.008.

(8) Moya, X.; Mathur, N. D. Caloric Materials for Cooling and Heating. *Science (1979)* **2020**, *370* (6518), 797–803. https://doi.org/10.1126/science.abb0973.

(9) Schipper, J.; Bach, D.; Mönch, S.; Molin, C.; Gebhardt, S.; Wöllenstein, J.; Schäfer-Welsen, O.; Vogel, C.; Langebach, R.; Bartholomé, K. On the Efficiency of Caloric Materials in Direct Comparison with Exergetic Grades of Compressors. *JPhys Energy* **2023**, *5* (4), 045002. https://doi.org/10.1088/2515-7655/ace7f4.

(10) Cirillo, L.; Greco, A.; Masselli, C. The Application of Barocaloric Solid-State Cooling in the Cold Food Chain for Carbon Footprint Reduction. *Energies (Basel)* **2023**, *16* (18), 6436. https://doi.org/10.3390/en16186436.

(11) Greco, A.; Aprea, C.; Maiorino, A.; Masselli, C. The Environmental Impact of a Caloric Heat Pump Working with Solid-State Materials Based on TEWI Analysis. *AIP Conf. Proc.* **2019**, 2191, 20091. https://doi.org/10.1063/1.5138824.

(12) Li, B.; Kawakita, Y.; Ohira-Kawamura, S.; Sugahara, T.; Wang, H.; Wang, J.; Chen, Y.; Kawaguchi, S. I.; Kawaguchi, S.; Ohara, K.; Li, K.; Yu, D.; Mole, R.; Hattori, T.; Kikuchi, T.; Yano, S. ichiro; Zhang, Z.; Zhang, Z.; Ren, W.; Lin, S.; Sakata, O.; Nakajima, K.; Zhang, Z. Colossal Barocaloric Effects in Plastic Crystals. *Nature* **2019**, *567* (7749), 506–510. https://doi.org/10.1038/s41586-019-1042-5.

(13) Aznar, A.; Lloveras, P.; Barrio, M.; Negrier, P.; Planes, A.; Mañosa, L.; Mathur, N. D.; Moya, X.; Tamarit, J.-L. L. Reversible and Irreversible Colossal Barocaloric Effects in Plastic Crystals. *J Mater Chem A Mater* **2020**, *8* (2), 639–647. https://doi.org/10.1039/c9ta10947a.

(14) Li, F.; Li, M.; Niu, C.; Wang, H. Atomic-Scale Insights into the Colossal Barocaloric Effects of Neopentyl Glycol Plastic Crystals. *Appl Phys Lett* **2022**, *120* (7), 073902. https://doi.org/10.1063/5.0081930.

(15) Somodi, C. B.; McCormick, K.; Tabor, D. P.; Pentzer, E.; Shamberger, P. J. Kinetics of the Plastic Crystal Transition in Neopentyl Glycol. *J Appl Phys* **2024**, *135* (14), 145101. https://doi.org/10.1063/5.0192791.

(16) Rendell-Bhatti, F.; Boldrin, D.; Dilshad, M.; Moya, X.; MacLaren, D. A. Understanding Variations of Thermal Hysteresis in Barocaloric Plastic Crystal Neopentyl Glycol Using Correlative Microscopy and Calorimetry. *JPhys Energy* **2024**, *6* (2), 025020. https://doi.org/10.1088/2515-7655/AD3985.

(17) Lilley, D.; Lau, J.; Dames, C.; Kaur, S.; Prasher, R. Impact of Size and Thermal Gradient on Supercooling of Phase Change Materials for Thermal Energy Storage. *Appl Energy* **2021**, *290*, 116635. https://doi.org/10.1016/j.apenergy.2021.116635.

(18) Li, F.; Niu, C.; Xu, X.; Li, M.; Wang, H. The Effect of Defect and Substitution on Barocaloric Performance of Neopentylglycol Plastic Crystals. *Appl Phys Lett* **2022**, *121* (22), 223902. https://doi.org/10.1063/5.0131123.

(19) Dai, Z.; She, X.; Shao, B.; Yin, E.; Ding, Y.; Li, Y.; Zhang, X.; Zhao, D. Plastic Crystal Neopentyl Glycol/Multiwall Carbon Nanotubes Composites for Highly Efficient Barocaloric Refrigeration System. *J. Therm. Sci.* **2024**, *33* (1), 383–393. https://doi.org/10.1007/s11630-023-1891-y.

(20) Chandra, D.; Ding, W.; Lynch, R. A.; Tomilinson, J. J. Phase Transitions in "Plastic Crystals." *J. Less-Common Met.* **1991**, *168* (1), 159–167. https://doi.org/10.1016/0022-5088(91)90042-3.

(21) Chandra, D.; Day, C. S.; Barrett, C. S. Low- and High-Temperature Structures of Neopentylglycol Plastic Crystal. *Powder Diffr* **1993**, *8* (2), 109–117. https://doi.org/10.1017/S0885715600017930.

(22) Andersson, M. S.; Grinderslev, J. B.; Jensen, T. R.; García Sakai, V.; Häussermann, U.; Udovic, T. J.; Karlsson, M. Interplay of $NH_4^+$ and $BH_4^-$ Reorientational Dynamics in $NH_4BH_4$. *Phys Rev Mater* **2020**, *4* (8), 085002. https://doi.org/10.1103/PhysRevMaterials.4.085002.

(23) Kruteva, M. Dynamics Studied by Quasielastic Neutron Scattering (QENS). *Adsorption*. Springer July 1, 2021, pp 875–889. https://doi.org/10.1007/s10450-020-00295-4.

(24) Li, F. B.; Li, M.; Xu, X.; Yang, Z. C.; Xu, H.; Jia, C. K.; Li, K.; He, J.; Li, B.; Wang, H. Understanding Colossal Barocaloric Effects in Plastic Crystals. *Nat Commun* **2020**, *11* (1), 1–8. https://doi.org/10.1038/s41467-020-18043-1.

(25) Bee, M. *Quasielastic Neutron Scattering, Principles and Applications in Solid State Chemistry, Biology and Materials Science*; Adam Hilger, 1988.

(26) Silvi, L.; Röhm, E.; Fichtner, M.; Petry, W.; Lohstroh, W. Hydrogen Dynamics in β-Mg(BH4)2 on the Picosecond Timescale. *Phys. Chem. Chem. Phys* **2016**, *18* (21), 14323–14332. https://doi.org/10.1039/c6cp00995f.

(27) Songvilay, M.; Wang, Z.; Sakai, V. G.; Guidi, T.; Bari, M.; Ye, Z. G.; Xu, G.; Brown, K. L.; Gehring, P. M.; Stock, C. Decoupled Molecular and Inorganic Framework Dynamics in $CH_3NH_3PbCl_3$. *Phys Rev Mater* **2019**, *3* (12), 125406. https://doi.org/10.1103/PhysRevMaterials.3.125406.

(28) Meijer, B. E.; Cai, G.; Demmel, F.; Walker, H. C.; Phillips, A. E. Pressure Dependence of Rotational Dynamics in Barocaloric Ammonium Sulfate. *Phys Rev B* **2022**, *106* (6), 064302. https://doi.org/10.1103/PhysRevB.106.064302.

(29) Hempelmann, R. *Quasielastic Neutron Scattering and Solid State Diffusion*; Oxford University Press, 2000. https://doi.org/10.1093/acprof:oso/9780198517436.001.0001.

(30) Sanuy, A.; Escorihuela-Sayalero, C.; Lloveras, P.; Tamarit, J. L.; Pardo, L. C.; Cazorla, C. Molecular Origins of Colossal Barocaloric Effects in Plastic Crystals. **2025**. https://doi.org/10.48550/arXiv.2501.14403.

(31) Seven, O.; Bolte, M. CCDC 832035: Experimental Crystal Structure Determination. 2011. 10.5517/ccwxsvd.

(32) Line, C. M. B.; Winkler, B.; Dove, M. T. Quasielastic Incoherent Neutron Scattering Study of the Rotational Dynamics of the Water Molecules in Analcime. *Phys Chem Miner* **1994**, *21* (7), 451–459. https://doi.org/10.1007/BF00202275.

(33) Lloveras, P.; Aznar, A.; Barrio, M.; Negrier, P.; Popescu, C.; Planes, A.; Mañosa, L.; Stern-Taulats, E.; Avramenko, A.; Mathur, N. D.; Moya, X.; Tamarit, J. L. Colossal Barocaloric Effects near Room Temperature in Plastic Crystals of Neopentylglycol. *Nat Commun* **2019**, *10* (1), 1–7. https://doi.org/10.1038/s41467-019-09730-9.

(34) Meijer, B. E.; Dixey, R. J. C.; Demmel, F.; Perry, R.; Walker, H. C.; Phillips, A. E. Dynamics in the Ordered and Disordered Phases of Barocaloric Adamantane. *Phys. Chem. Chem. Phys* **2023**, *25*, 9282. https://doi.org/10.1039/d2cp05412d.

(35) Font, J.; Muntasell, J.; Navarro, J.; Tamarit, J. L.; Lloveras, J. Calorimetric Study of the Mixtures PE/NPG and PG/NPG. *Sol. Energy Mater.* **1987**, *15* (4), 299–310. https://doi.org/10.1016/0165-1633(87)90045-1.

(36) Tamarit, J. L.; Legendre, B.; Buisine, J. M. Thermodynamic Study of Some Neopentane Derivated by Thermobarometric Analysis. *Molecular Crystals and Liquid Crystals Science and Technology. Section A. Molecular Crystals and Liquid Crystals* **1994**, *250* (1), 347–358. https://doi.org/10.1080/10587259408028219.

(37) Kamae, R.; Suenaga, K.; Matsuo, T.; Suga, H. Low-Temperature Thermal Properties of 2,2-Dimethyl-1,3-Propanediol and Its Deuterated Analogues. *J. Chem. Thermodyn* **2001**, *33* (5), 471–484. https://doi.org/10.1006/jcht.2000.0694.

(38) Frick, B.; Combet, J.; Van Eijck, L. New Possibilities with Inelastic Fixed Window Scans and Linear Motor Doppler Drives on High Resolution Neutron Backscattering Spectrometers. *Nucl Instrum Methods Phys Res A* **2012**, *669*, 7–13. https://doi.org/10.1016/j.nima.2011.11.090.

(39) Lerbret, A.; Bordat, P.; Affouard, F.; Guinet, Y.; Hédoux, A.; Paccou, L.; Prévost, D.; Descamps, M. Influence of Homologous Disaccharides on the Hydrogen-Bond Network of Water: Complementary Raman Scattering Experiments and Molecular Dynamics Simulations. *Carbohydr Res* **2005**, *340* (5), 881–887. https://doi.org/10.1016/j.carres.2005.01.036.

(40) Crupi, V.; Longo, F.; Majolino, D.; Venuti, V. The Hydrogen-Bond Network in Propylene-Glycol Studied by Raman Spectroscopy. *J Mol Struct* **2006**, *790* (1–3), 141–146. https://doi.org/10.1016/j.molstruc.2005.10.052.

(41)  Lu, W.; Amarasinghe, C.; Zhang, E.; Martin, A.; Kaur, S.; Prasher, R.; Ahmed, M. Probing Hydrogen-Bond Networks in Plastic Crystals with Terahertz and Infrared Spectroscopy. *Cell Rep Phys Sci* **2022**, *3* (8), 100988. https://doi.org/10.1016/j.xcrp.2022.100988.

(42)  Aznar, A.; Negrier, P.; Planes, A.; Mañosa, L.; Stern-Taulats, E.; Moya, X.; Barrio, M.; Tamarit, J. L.; Lloveras, P. Reversible Colossal Barocaloric Effects near Room Temperature in 1-X-Adamantane (X=Cl, Br) Plastic Crystals. *Appl Mater Today* **2021**, *23*, 101023. https://doi.org/10.1016/j.apmt.2021.101023.

(43)  Salvatori, A.; Aguilà, D.; Aromí, G.; Mañosa, L.; Planes, A.; Lloveras, P.; Pardo, L. C.; Appel, M.; Nataf, G. F.; Giovannelli, F.; Barrio, M.; Tamarit, J. L.; Romanini, M. Large Barocaloric Effects in Two Novel Ferroelectric Molecular Plastic Crystals. *J Mater Chem A Mater* **2023**, *11*, 12140–12150. https://doi.org/10.1039/d2ta10033a.

(44)  Vispa, A.; Monserrat, D.; Cuello, G. J.; Fernandez-Alonso, F.; Mukhopadhyay, S.; Demmel, F.; Tamarit, J. L.; Pardo, L. C. On the Microscopic Mechanism behind the Purely Orientational Disorder-Disorder Transition in the Plastic Phase of 1-Chloroadamantane. *Phys. Chem. Chem. Phys* **2017**, *19* (30), 20259–20266. https://doi.org/10.1039/c7cp03630b.